\documentclass[prd,twocolumn,floats,floatfix,nofootinbib]{revtex4}
\usepackage[dvips]{graphicx}
\usepackage{graphics}
\usepackage{dcolumn}
\usepackage{amssymb}
\usepackage{bm}
\usepackage{amsmath}

\def\spose#1{\hbox to 0pt{#1\hss}}

\def\lta{\mathrel{\spose{\lower 3pt\hbox{$\mathchar"218$}}
     \raise 2.0pt\hbox{$\mathchar"13C$}}}
\def\gta{\mathrel{\spose{\lower 3pt\hbox{$\mathchar"218$}}
     \raise 2.0pt\hbox{$\mathchar"13E$}}}
\newcommand{\be}{\begin{equation}}
\newcommand{\en}{\end{equation}}
\newcommand{\bea}{\begin{eqnarray}}
\newcommand{\ena}{\end{eqnarray}}

\begin{document}

\title[ ]{Gravitational wave momentum extraction in non-axisymmetric Robinson-Trautman spacetimes}

\author{R. F. Aranha$^{1}$, I. Dami\~ao Soares$^{1}$ and E. V. Tonini$^{2}$}

\address{$^{1}$Centro Brasileiro de Pesquisas F\'isicas, Rio de Janeiro 22290-180, Brazil,\\
$^{3}$Instituto Federal do Esp\'irito Santo, Vit\'oria 29040-780, Brazil.}
\email{rfaranha@cbpf.br; ivano@cbpf.br;tonini@cefetes.br }

\date{\today}
\begin{abstract}
We examine numerically the gravitational wave recoil in non-axisymmetric Robinson-Trautman spacetimes.
We construct characteristic initial data for the Robinson-Trautman dynamics which are interpreted
as corresponding to the early post-merger state of two boosted colliding black holes with a
common apparent horizon. Our analysis is based on the Bondi-Sachs energy-momentum conservation laws
which regulate the radiative transfer processes involved in the emission of gravitational
waves. We evaluate the Bondi-Sachs momentum flux carried out by gravitational waves
and the associated net kick velocity defined (in a zero-initial-Bondi-momentum frame) as proportional
to the total gravitational wave impulse imparted on the system.
The kick velocity distributions are obtained and analyzed for two distinct classes of
initial data corresponding to the early post-merger state of
(i) non-head-on collisions and (ii) head-on collisions of black holes. For the first class (i), the net gravitational
wave momentum fluxes and associated kicks are evaluated for a given domain of parameters (incidence angle and mass ratio).
Typically for the equal mass case the net gravitational wave momentum flux carried is nonzero.
This last result indicates that these configurations are not connected with black hole binary inspirals or
head-on collisions. We suggest that these systems might be a candidate to an approximate description
of the post-merger phase of a non-head-on collision of black holes not preceded by a pre-merger
inspiral phase, as for instance colliding black holes in pre-merger unbounded trajectories.
For the second class (ii), head-on collisions, we compare our results, and discuss the analogies,
with $1+3$ numerical relativity simulations of binary black hole inspirals and head-on collisions.
\end{abstract}


\maketitle
{\small
\section{Introduction\label{sec1}}

The collision and merger of two black holes is presently considered to be an important astrophysical
configuration where processes of generation and emission of gravitational waves take place (cf. \cite{pretorius}
and references therein). The radiative transfer involved in these processes, evaluated in the full nonlinear
regime of General Relativity, shows that gravitational waves extract mass, momentum and angular momentum of
the source, and may turn out to be fundamental for the astrophysics of the collapse of stars and the formation
of black holes. The process of momentum extraction and the associated recoils
in the system can have important consequences for astrophysical scenarios, as the evolution and the
population of massive black holes in galaxies or in the intergalactic medium\cite{baker,merritt,favata}.
Observational evidence of black hole recoils have been reported in \cite{komossa} and references therein.
\par The kick processes and associated recoils in the merger of two black holes have been investigated within
several approaches, most of them connected to binary black hole inspirals.
Post-Newtonian approximations (cf. \cite{blanchet1} and references therein) estimated the
kick velocity accumulated during the adiabatical inspiral of the system
plus the kick velocity accumulated during the plunge phase.
Sopuerta et al.\cite{sopuerta} computed the recoil velocity based on the close limit approximation (CLA)
supplemented with post-Newtonian (PN) calculations, and estimated a lower bound for the kick velocity
in binaries with a mass ratio $\simeq 0.34$.
The first full numerical relativity evaluation of the recoil in nonspinning black hole binaries
was reported by Baker et al.\cite{baker} for a mass ratio $\simeq 0.667$ of the two black holes, while
Gonz\'alez et al.\cite{gonzalez0} and Campanelli et al.\cite{camp1} simultaneously obtained much larger
recoils for black hole binaries with antialigned spins.
Gonz\'alez et al.\cite{gonzalez} undertook a more complete full numerical relativity examination
of kicks in the merger of nonspinning black hole binaries by contemplating a larger parameter domain.
For the case of small mass ratios in the interval $0.01 \leq \alpha \leq 0.1$ full numerical relativity
evaluations bridged with perturbative techniques were implemented in Refs. \cite{gonzalez1,camp3,camp4,camp2}.
Le Tiec et al.\cite{blanchet}, combining PN+CLA methods, recently evaluated the gravitational wave recoil
in black hole binaries and showed that the ringdown (evaluated within the CLA) produces a significant anti-kick.
In the same vein Choi et al.\cite{choi} examined recoils in head-on collisions of spinning and nonspinning
black holes, considering the head-on case as a model problem which can be seen as an approximation to the final plunge to merger
and allow to isolate kick effects from the orbital inspiral motion.
Finally Rezzola et al.\cite{rezzolla,rezzolla1} obtained an important injective relation between
the kick velocities and the effective curvature parameter of the global apparent horizon, in head-on collisions,
using initial data derived in \cite{aranha1,aranhaT}.
\par
From numerical relativity $3+1$ evaluations we have evidence that an important contribution to the
gravitational wave recoil comes from the post-merger phase, where also a final deceleration regime is present
leading to the so termed anti-kicks in the system. In the present paper we intend to approach
some of these issues by examining the gravitational wave recoil in non-axisymmetric
Robinson-Trautman (RT) spacetimes\cite{rt} in the Bondi-Sachs (BS) characteristic formulation of
gravitational wave emission\cite{ bondi,sachs}. The paper extends and
completes Ref. \cite{aranha11} where we examined numerically the gravitational
wave production and related radiative processes using initial data
that are interpreted as describing the post-merger phase of non-head-on colliding black holes.
Our numerical codes are accurate and sufficiently stable for long time runs so that
we are able to reach numerically the final configuration of the system, when the gravitational emission ceases.
RT spacetimes already present an apparent horizon so that the dynamics covers the post merger
phase of the system up to the final configuration of the remnant black hole.
Due to the presence of a global apparent horizon the initial data  effectively
represents an initial single distorted black hole which is evolved via the RT dynamics.
Similarly to the case of the CLA -- where the perturbation equations of a black hole\cite{zerilli,teuk}
are feeded either with numerically generated, or with Misner, or Bowen-York-type initial data --
we feed the (nonlinear) RT equation with the above mentioned characteristic data.
It is in this sense that we denote the dynamics thus generated as ``the post-merger phase of
two colliding black holes''. The interpretation of the outcomes of the RT dynamics
and its comparison, in the head-on case, with numerical relativity results should
be considered with the above caveats.
\par RT spacetimes\cite{rt} are asymptotically flat solutions of Einstein's vacuum equations
that describe the exterior gravitational field of a bounded system
radiating gravitational waves. In a suitable coordinate system the metric can be expressed as
{\small
\begin{eqnarray}
\label{eq1}
\nonumber
ds^2&=&\Big({\lambda(u,\theta,\phi)- \frac{2 m_{0}}{r}+ 2 r \frac{{K}_{,u}}{K}}\Big) d u^2 +2du dr\\
&-&r^{2}K^{2}(u,\theta,\phi)~ \Big( d \theta^{2}+\sin^{2}\theta d \phi^{2} \Big),
\end{eqnarray}}
where
{\small
\begin{eqnarray}
\lambda(u,\theta,\phi)=\frac{1}{K^2}-\frac{(K_{,\theta}~ \sin \theta/K)_{,\theta}}{K^2 \sin \theta}
+\frac{1}{\sin^2 \theta}\Big(\frac{K_{,\phi}^2}{K^4}-\frac{K_{,\phi \phi}}{K^3} \Big).
\label{eq2}
\end{eqnarray}}
The Einstein vacuum equations for (\ref{eq1}) result in
{\small
\begin{eqnarray}
\label{eq3}
-6 m_{0}\frac{{K}_{,u}}{K}+\frac{1}{2 K^2}\Big(\frac{(\lambda_{,\theta} \sin
\theta)_{,\theta}}{ \sin \theta}+\frac{\lambda_{,\phi \phi}}{ \sin^2 \theta}\Big)=0.
\end{eqnarray}}
\noindent Subscripts $u$, $\theta$ and $\phi$, preceded by a comma, denote derivatives with respect to
$u$, $\theta$ and $\phi$, respectively. $m_0 > 0$ is the only dimensional parameter of the geometry, which
fixes the mass and length scales of the spacetime. Eq. (\ref{eq3}), the RT equation, governs the dynamics
of the system and allows to evolve the initial data $K(u_0,\theta,\phi)$, given in the characteristic surface $u=u_0$,
for times $u > u_0$. For sufficiently regular initial data RT spacetimes exist globally for all positive $u$
and converge asymptotically to the Schwarzschild metric as $u \rightarrow \infty$\cite{chrusciel}.
Once the initial data $K(u_0,\theta,\phi)$ is specified, a unique apparent horizon (AH) solution
is fixed for that $u_0$\cite{AH}. Since the AH is the outer past marginally trapped surface,
the closest of a white hole definition (the remnant black hole will form as $u \rightarrow \infty$),
only the exterior and its future development via RT dynamics with outgoing gravitational waves is of interest.
We note that all the BS quantities, measured at the future null infinity $\mathcal{J}^{+}$,
are constructed and well defined under the outgoing radiation condition[17,18].
\par The field equations have a stationary solution which will play an important role in our discussions,
{\small
\begin{equation}
K(\theta,\phi)=\frac{K_0}{\cosh \gamma+ ({\bf{n}} \cdot \hat{\bf{x}})\sinh \gamma},
\label{eq4}
\end{equation}}

\noindent where $\hat{\bf{x}}=(\sin \theta \cos \phi,\sin \theta \sin \phi, \cos \theta)$ is the unit
vector along an arbitrary direction ${\bf{x}}$, and ${\bf{n}}=(n_1,n_2,n_3)$ is a constant unit vector
satisfying $n_{1}^{2}+n_{2}^{2}+n_{3}^{2}=1$. The $K_0$ and $\gamma$ are constants. We note
that (\ref{eq4}) yields $\lambda=1/K_{0}^{2}$, resulting in its stationary character (cf. eq. (\ref{eq3})).
This solution can be interpreted\cite{bondi} as a boosted Schwarzschild black hole along the axis determined
by ${\bf{n}}$, with boost parameter $\gamma$ (or velocity parameter $v=\tanh \gamma$). For $\gamma=0$ we recover
the Schwarzschild black hole at rest.
\par The Bondi mass function of (\ref{eq4}) is $m(\theta,\phi)=m_{0}K^3(\theta,\phi)$, and the total
mass-energy of this gravitational configuration is given by the Bondi mass
{\small
\begin{eqnarray}
\label{eq5}
\nonumber
M_B&=&\frac{1}{4 \pi}\int^{2\pi}_{0} d \phi \int^{\pi}_{0}  m(\theta,\phi)\sin \theta ~d\theta\\
&=&m_{0} K_{0}^{3}\cosh \gamma = m_{0} K_{0}^{3}/ \sqrt{1-v^2}.
\end{eqnarray}}
The interpretation of (\ref{eq4}) as a boosted black hole is relative to
the asymptotic Lorentz frame which is the rest frame of the black hole when $\gamma=0$.
\par In the paper we use geometrical units $G=c=1$; $c$ is however restored in the
definition of the kick velocity. Except where explicitly stated, all the numerical
results are for $\gamma=0.5$ fixed. In our computations we adopted $m_0=10$
but the results are given in terms of $u/m_0$. We note that we can always set $m_0=1$ in RT equation
(\ref{eq3}) by the transformation $u \rightarrow {\tilde{u}}=u/m_0$.
\section{The Bondi-Sachs four momentum for RT spacetimes, the news and the initial data\label{sec2}}
\par Since RT spacetimes describe asymptotically flat radiating spacetimes and the initial data
for the dynamics are prescribed on null characteristic surfaces, they are in the realm of
the $2+2$ Bondi-Sachs (BS) formulation of gravitational waves in General Relativity\cite{bondi,sachs,sachs1}.
Therefore we shall use the physical quantities of this formulation which appear in the description of
gravitational wave emission processes, as the BS four-momentum and its conservation laws.
A detailed derivation of the BS four-momentum conservation laws in RT spacetimes was given
in \cite{cqgIvano} but, for the sake of clarity, a few fundamental results are discussed in the following.
From the supplementary vacuum Einstein equations $R_{UU}=0$, $R_{U \Theta}=0$, and $R_{U \Phi}=0$ in the
BS formulation (where $(U,R,\Theta,\Phi)$ are Bondi-Sachs coordinates),
we obtain\cite{bondi,sachs}
{\small
\begin{eqnarray}
\label{eq6}
\nonumber
&&\frac{\partial m(u,\theta,\phi)}{\partial u}= - K \Big( {c_{,u}^{(1)}}^2 + {c_{,u}^{(2)}}^2 \Big)
+ \frac{1}{2}\frac{\partial}{\partial u}
\Big[3 c_{,\theta}^{(1)} \cot \theta \\&+& 4 c_{,\phi}^{(2)} \frac{\cos \theta}{\sin^2 \theta} -2 c^{(1)} + c_{,\theta \theta}^{(1)}+ \frac{2}{\sin \theta}
 c_{,\theta \phi}^{(2)} - \frac{1}{\sin^2 \theta} c_{,\phi \phi}^{(1)} \Big],
\end{eqnarray}}
where $m(u,\theta,\phi)$ is the Bondi mass function and $c_{,u}^{(1)}(u,\theta,\phi)$ and $c_{,u}^{(2)}(u,\theta,\phi)$ are the two
{\it news} functions, in RT coordinates. The extra factor $K$ in the first term of the right-hand-side of Eq. (\ref{eq6})
comes from the transformation of the Bondi time coordinate $U$ to the RT coordinate
$u$, ${\rm lim}_{r \rightarrow \infty}=\partial U/\partial u=1/K$.
For the RT spacetimes (\ref{eq1}), the {\it news} are expressed as\cite{cqgIvano}
{\small
\begin{eqnarray}
\label{eq7}
\nonumber
c_{,u}^{(1)}(u,\theta,\phi)&=&\frac{1}{2} \Big( {\mathcal P}_{,\theta \theta}-{\mathcal P}_{,\theta} \cot \theta -\frac{{\mathcal P}_{,\phi\phi}}{\sin^2 \theta}\Big),\\
c_{,u}^{(2)}(u,\theta,\phi)&=&\frac{1}{\sin \theta} \Big({\mathcal P}_{,\theta \phi}- {\mathcal P}_{,\phi} \cot \theta\Big),
\end{eqnarray}}
where we have introduced the variable ${\mathcal P} \equiv 1/K$, for notation convenience.
We remark that $c^{(1)}_{,u}=0=c^{(2)}_{,u}$ for the boosted Schwarzschild solution (\ref{eq4}), as should be expected.
\par The Bondi-Sachs four momentum is defined as\cite{sachs1}
{\small
\begin{eqnarray}
\label{eq9}
P^{\mu}(u)= \frac{1}{4 \pi} \int^{2 \pi}_{0} d \phi \int^{\pi}_{0} m(u,\theta,\phi)~l^{\mu} \sin \theta~ d \theta,
\end{eqnarray}}
\noindent where $l^{\mu}=(1, -\sin \theta \cos \phi,-\sin \theta \sin \phi, -\cos \theta)$, relative to an
asymptotic Lorentz frame\footnote{The four vector $l^{\mu}$ actually defines the generators $l^{\mu} \Big(\partial/\partial U \Big)$ of the BMS translations in the
temporal and Cartesian $x,y,z$ directions of the asymptotic Lorentz frame\cite{sachs1}.}. From Eq. (\ref{eq6}) the Bondi-Sachs four momentum conservation law follows
{\small
\begin{eqnarray}
\label{eq12}
\frac{d P^{\mu}(u)}{d u}= P_{W}^{\mu}(u),
\end{eqnarray}}
where
{\small
\begin{eqnarray}
\label{eq11}
P_{W}^{\mu}(u)= -\frac{1}{4 \pi} \int^{2 \pi}_{0} d \phi \int^{\pi}_{0} K~l^{\mu} \Big( {c_{,u}^{(1)}}^2 + {c_{,u}^{(2)}}^2 \Big) \sin \theta~ d \theta~~
\end{eqnarray}}
is the net flux of energy-momentum carried out by the the gravitational waves emitted.
We note that the term between square brackets in the right-hand-side of (\ref{eq6}) vanishes in the integrations
due to the boundary conditions satisfied by the {\it news}, $c^{(1)}=c^{(2)}=0$ and
$c^{(1)}_{\theta}=c^{(2)}_{\theta}=0$ at $\theta=0, \pi$.
The mass-energy conservation law (Eq. (\ref{eq12}) for $\mu=0$) is the Bondi mass formula.
Our main interest here is the analysis of the linear momentum conservation, Eq. (\ref{eq12}) for $\mu=x,y,z$,
{\small
\begin{eqnarray}
\label{eq13}
\frac{d{\bf P}(u)}{d u}={\bf P}_{W}(u),
\end{eqnarray}}
where ${\bf P}_W(u)=(P^{x}_{W}(u),P^{y}_{W}(u),P^{z}_{W}(u))$, with
{\small
\begin{eqnarray}
\label{eq15-i}
{P}_{W}^{x}(u)&=&\frac{1}{4 \pi}  \int^{2 \pi}_{0}  d \phi \int^{\pi}_{0} \sin^2 \theta \cos \phi~ K \Big( {c_{,u}^{(1)}}^2 + {c_{,u}^{(2)}}^2 \Big) d \theta,~~~~~~\\
\label{eq15-ii}
{P}_{W}^{y}(u)&=&\frac{1}{4 \pi}  \int^{2 \pi}_{0} d \phi \int^{\pi}_{0} \sin^2 \theta \sin \phi~ K \Big( {c_{,u}^{(1)}}^2 + {c_{,u}^{(2)}}^2 \Big) d \theta,~~~~\\
\label{eq15-iii}
{P}_{W}^{z}(u)&=&\frac{1}{4 \pi}  \int^{2 \pi}_{0} d \phi \int^{\pi}_{0} \cos \theta \sin \theta~ K \Big( {c_{,u}^{(1)}}^2 + {c_{,u}^{(2)}}^2 \Big) d \theta.~~~~
\end{eqnarray}}
\begin{figure}
\begin{center}
{\includegraphics*[height=3.9cm,width=8.2cm]{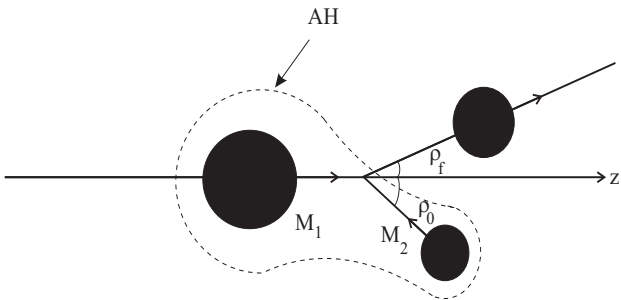}}
\caption{Schematic diagram illustrating the initial data (\ref{eq16}) corresponding to the non-head-on collision of two
black holes, reproduced from \cite{aranha11}. The dashed line depicts the common apparent horizon (AH). The remnant black hole
is also displayed, where $\rho_f$ defines its direction of motion. The $x$-axis is upward.}
\label{figDD}
\end{center}
\end{figure}
\par The characteristic initial data to be used was derived in \cite{aranha11} and can be interpreted as
describing the early post-merger configuration of two colliding black holes in the $(x,z)$ plane, at $u=u_0$,
{\small
\begin{eqnarray}
\nonumber
&&K(u_0,\theta,\phi)=\Big(\frac{\alpha_1}{\sqrt{{\cosh \gamma+ \cos \theta \sinh \gamma}}}+\\
&&\frac{\alpha_2}{\sqrt{{\cosh \gamma - (\cos \rho_0 ~\cos \theta- \sin \rho_0~ \sin\theta \cos \phi)\sinh \gamma}}}\Big)^2.~~~~~
\label{eq16}
\end{eqnarray}}
In the derivation of (\ref{eq16}) it turns out that $\alpha_2/\alpha_1$ is the mass ratio of the
Schwarzschild masses of the initial data, as seen by an asymptotic observer. We also mention the following
properties of (\ref{eq16}): (i) for $\alpha_2=0$ ($\alpha_1 \neq 0$) or $\alpha_1=0$ ($\alpha_2 \neq 0$)
the initial data (\ref{eq16}) corresponds to a
boosted Schwarzschild black hole along, respectively, the positive $z$-axis or along the direction
of the unit vector ${\bf n}=(-\cos \rho_0,0,\sin \rho_0)$, with respect to an asymptotic Lorentz
frame (cf. eq. (\ref{eq4})). (ii) The specific combination (\ref{eq16}) of two boosted black hole solutions (\ref{eq4})
is not arbitrary but arises as the conformal factor of an asymptotically flat 3-geometry
which is a solution of the constraint $^{(3)} R=0$.
(iii) Additionally (\ref{eq16}) results in a planar dynamics, namely, for all $u$ the net gravitational wave
momentum flux ${\bf P}_W(u)$ is restricted to the plane determined
by the unit vectors ${\bf n}_z$ and ${\bf n}$ defining the direction of motion of the two black holes entering in (\ref{eq16}),
namely, the $(x,z)$-plane. Therefore we have that $P_{W}^{y}(u)=0$ for all $u$, implying obviously that $ P^{y}(u)$ is conserved.
The momentum of the remnant is also contained in this plane, as should be expected.
(iv) The data (\ref{eq16}) with $\rho_0=0^{\circ}$ is the only combination of the black hole solutions (\ref{eq4})
that results in a distribution for the recoils satisfying a Fitchett law, as shown in the many numerical tests we have performed.
The above properties are consistent with the interpretation of (\ref{eq16}) as related to the early post-merger state of a black hole collision.
\par As concerning the parameter $\rho_0$, the data corresponds to bring instantaneously in interaction the two individual boosted black holes,
each described by equation (\ref{eq4}), one along the $z$ axis (${\bf{n}}=(0,0,1)$) and the other with $\bf{n}$ making an angle $\rho_0$
with the $z$-axis of the asymptotic Lorentz observer. $\gamma$ is the boost parameter of the black hole solutions (\ref{eq4})
which enter in (\ref{eq16}).
As in the characteristic and $1+3$ numerical relativity approaches, the interpretation
of the initial data parameters involves an approximation, namely, that the initial gravitational
interaction in the data is neglected.
A schematic diagram of the initial data is given in Fig. \ref{figDD}, reproduced from \cite{aranha11}.
As mentioned already this data has a common apparent horizon so that the evolution covers the post merger
regime up to the final configuration, when the gravitational wave emission ceases.
In the remaining of the paper we fix $\alpha_1=1$ and relabel $\alpha_2 \equiv \alpha$.
\par Finally we must address the issue of {\it junk} radiation as concerning our initial data.
In fact, on evolving the system, we did not obtain evidence of
any {\it junk} radiation component present in the initial data. The time behavior of the radiated
energy and momentum fluxes, from the initial configuration up to the final configuration, is very smooth,
with no transient component or ``bump'' in its early evolution. This can be seen, for instance, in the
momentum fluxes of Figs. \ref{fig1} and \ref{fig3}(top) of Section \ref{section4}, where the
initial domain of the time axes is stretched by the use of a log scale. This pattern of the fluxes
is stable for variations of the parameters of the initial data. Similarly the gravitational wave energy fluxes
(examined in \cite{aranha11}) decrease monotonically with $u$ for any domain of parameters.
Therefore the result of a non-zero gravitational wave recoil for $\alpha=1$ and $\rho_0 \neq 0$
reported in Section \ref{section4} is unlikely to be a contribution of a junk radiation component.

\section{Numerical evolution\label{section3}}
The initial data (\ref{eq16}) is evolved numerically via the RT equation (\ref{eq3}), which is integrated using a
Galerkin spectral method with a spherical harmonics projection basis space\cite{fletcher}
adapted to the non-axisymmetric dynamics of RT spacetimes. The implementation of the Garlenkin method
is described in detail in Section V of Ref. \cite{aranha11}, as well as its accuracy and stability for
long time runs. However for completeness we present here a brief survey of the method.
{\small
\begin{center}
\begin{table*}
\caption{$\chi_1(u) \equiv \big(A_{N_P,0}/A_{0,0}\Big)(u)$, $\chi_2 (u)\equiv \Big(A_{N_P,N_P}/A_{0,0}\Big)(u)$, $u_f \simeq 460$.}
{\small
\begin{tabular}{|c|c|c|c|c|}
\hline
$ $ & $\chi_1(0)$,~~$\chi_2(0)$ & $\chi_1(110)$,~~$\chi_2(110)$ & $\chi_1(250)$,~ $\chi_2(250)$ & $\chi_{1,2}(u_f)$    \\ \hline
$N_P=5$ & $$1.0~{\rm e-4}$$,~ $$2.8~{\rm e-10}$$ &$$1.3{\rm e-6}$$,~$$ 0 $$ & $$0.8~{\rm e-9}$$,~$$ 0 $$& $0$  \\ \hline
$N_P=6$ & $$-9.2~{\rm e-6}$$,~$$-1.9~{\rm e-12}$$ &$$-2.5~{\rm e-7}$$,~$$0~ $$  & $$-1.6~{\rm e-10}$$, $$0~ $$ &$0$  \\ \hline
$N_P=7$ & $$2.0~{\rm e-6}$$,~$$1.0~{\rm e-14}$$  & $$4.3~{\rm e-8}$$,~$$0 $$ &  $$2.7~{\rm e-11}$$, $$0 $$ & $0$ \\ \hline
\end{tabular}}
\label{table1}
\end{table*}
\end{center}}
\par For convenience we will use the function ${\mathcal P}(u,\theta,\phi) \equiv 1/K(u,\theta,\phi)$.
The Galerkin method establishes that ${\mathcal P}(u,\theta,\phi)$ can be expanded in a convenient set of basis
functions of a projection space (in this case the real spherical harmonics basis) through which we can reduce
the RT partial differential equation into an autonomous nonlinear dynamical system for the modal
coefficients $(A_{l,m}(u),B_{l,m}(u))$ of the expansion ${\mathcal P}(u,\theta,\phi)= \sum_{l=1}^{N_{P}} \sum_{m=1}^{l} A_{l,m}(u) P_{l}^{m}(\cos \theta) \cos m \phi+\sum_{l=1}^{N_{P}} \sum_{m=1}^{l} B_{l,m}(u) P_{l}^{m}(\cos \theta) \sin m \phi$, where $P_{l}^{m}(\cos \theta)$ are the associated Legendre functions properly normalized. Projecting this expression on each element of the
basis of the projection space (using the ${\mathcal{L}}^2$ norm and the orthonormal relations of the basis under this norm)
we obtain the modal coefficients $A_{l,m}(u)$ , $B_{l,m}(u)$ for the corresponding ${\mathcal P}(u,\theta,\phi)$. $N_P$ is the truncation of
the method and, together with the choice of the projection space, constitutes a fundamental cornerstone of the method.
Substituting this expansion, and its time and angle derivatives, in the RT equation (\ref{eq3}) and projecting on each
element of the basis we obtain an autonomous dynamical system of dimension $(N_P + 1)^2$ for the modal coefficients, ${\dot A}_{l,m}(u)= {\mathcal{A}}_{l,m} \Big( A_{l,m}(u),B_{l,m}(u) \Big)$, ${\dot B}_{l,m}(u)= {\mathcal{B}}_{l,m} \Big( A_{l,m}(u),B_{l,m}(u) \Big)$, where ${\mathcal{A}}_{l,m}$ and ${\mathcal{B}}_{l,m}$ are polynomials of order $(N_P +1)^2$ in $A_{l,m}(u)$ and $B_{l,m}(u)$. Contrary to usual numerical methods, whose objective is to obtain an
approximate solution of the exact dynamics, the Galerkin method in essence has the task of obtaining the solution of an approximate
dynamics described by the above dynamical system. The approximation consists in choosing a suitable finite $N_P$ that makes the method
sufficiently accurate and saves computational effort. Of course the exact dynamics corresponds to $N_P=\infty$.
In the integration of the autonomous dynamical system for the modal coefficients we used a fourth-order Runge-Kutta
recursive method (adapted to our constraints) together with a C++ integrator\cite{aranha11}.
For the numerical calculations in this paper we adopted $N_P=7$, which approximates quite well not
only the initial data but also the dynamics, as we discuss now. In the above paragraph the $u$-derivative
was denoted by a superscript dot to avoid the overcluttering of indices in the formulas.
\par A basic numerical test for the accuracy of the dynamics is connected to the planar nature of the collision
for initial data in which the initial directions of motion $({\bf n}_z, {\bf n})$ are restricted to an arbitrary
plane\cite{aranha11}. The approximated dynamics of the modal coefficients $(A_{l,m}(u),B_{l,m}(u))$ with $N_P=7$ preserves accurately
the planar motion which has led us to the simplified form (\ref{eq16}), corresponding to initial motion in the $x-z$ plane.
In this case the modal coefficients $B_{l,m}(u)$ are zero for all $u$ (within machine precision, meaning $\leq 10^{-14}$)
and a lot of computational effort is saved by reducing the Galerkin expansion to the $\cos m \phi$ series, with the dynamics
restricted to the modal coefficients $A_{l,m}$ only. In this instance the Bondi-Sachs momentum in the $y$-direction must
be conserved and the net momentum flux $P_{W}^{y}(u)=0$ for all $u$. This is accurately maintained by the numerical evolution
of the modal coefficients $A_{l,m}(u)$ with truncation $N_P=7$, namely, the reconstructed net momentum flux
$P_{W}^{y}(u)=0$ for all $u$, within machine precision.
\par Also the power in the highest modes $l=7$ is relatively very small at all times as compared to
the lowest dynamical mode $A_{0,0}$, so that the convergence of the method is guaranteed.
We illustrate this with the initial data for $\alpha=0.2$, $\rho_0=21^{\circ}$ and $\gamma=0.5$.
In all our computations we adopted $m_0=10$. In Table \ref{table1} we give the ratio of the highest modes, $\chi_1(u) \equiv A_{N_P,0}/A_{0,0}$
and $\chi_2 (u)\equiv A_{N_P,N_P}/A_{0,0}$ for truncations $N_P=5$, $N_P=6$ and $N_P=7$, sampled for several times from $u=0$
to the final time of computation $u=u_f \simeq 460$, when all modal coefficients are zero (within machine precision)
except $A_{0,0}$, $A_{0,1}$ and $A_{1,1}$. These latter nonzero modal coefficients characterize the remnant black hole.
{\small
\begin{center}
\begin{table*}
\caption{The net kick velocity $V_k$ ({\rm km/s}) for several values of $\alpha$ and
truncations $N_P=5$, $N_P=6$ and $N_P=7$.}
\vspace{0.15cm}
{\small
\begin{tabular}{|c|c|c|c|c|}
\hline
$ $ & ~~$\alpha=0.1$ ~~& ~~$\alpha=0.3$~~ & ~~$\alpha=0.5$ ~~& ~~$\alpha=0.9$~~    \\ \hline
$N_P=5$ & 25.068 & 97.036 & 120.288 & 108.474  \\ \hline
$N_P=6$ & 25.231 &97.298 & 120.430 & 108.448 \\ \hline
$N_P=7$ & 25.251  & 97.327 & 120.430 & 108.455 \\ \hline
\end{tabular}}
\label{table2}
\end{table*}
\end{center}}
\par The corresponding integrated error in the approximation of the initial data,
{\small $R(u_0,\phi)=\int_{0}^{\pi} {\mathcal R}(u_0,\theta,\phi) \sin \theta d \theta$} (where {\small ${\mathcal R}(u_0, \theta, \phi)=|{\mathcal P}(u_0,\theta,\phi)-{\mathcal P}_{\rm appr}(u_0,\theta,\phi)/{\mathcal P}(u_0,\theta,\phi)|$}) is $\leq 8.3 \times 10^{-6}$ for $N_P=6$,
and $\leq 1.6 \times 10^{-6}$ for $N_P=7$ for all $\phi$. These orders of magnitude are  maintained for the domain of $\alpha$
considered, and up to $\gamma=0.7$. The smaller error $\leq 6.7 \times 10^{-10}$ reported in \cite{aranha11} is related to a boost parameter $\gamma=0.2$.
The maximum relative errors {\small ${\mathcal R}(u_0, \theta,\phi)$} have orders of magnitudes $\sim 10^{-5}$ for $N_P=5$ and
$\sim 10^{-6}$ for $N_P=6$ and $N_P=7$, but for a substantial domain of $(\theta,\phi)$ these errors can be extremely small.
\par The best indicator to the accuracy of the expansion of the initial data, and of the RT dynamics for a given truncation $N_P$, is the behavior of
the constant of motion present in RT dynamics, $\zeta(u)=(1/4\pi) \int^{2\pi}_{0} d \phi \int^{\pi}_{0} K^{2}(u,\theta,\phi) \rm sin \theta d \theta$.
Its exact value can be obtained using the exact initial data which, for the present case $\alpha=0.2,~\gamma=0.5,~\rho_0=21^{\circ}$, is
$\zeta_{\rm exact}=1.94727076130606$. For the approximated initial data, with truncation $N_P=7$, we obtain the value
$\zeta_{\rm appr}(u=0)=1.94727076160249$ with an error $\Delta \zeta(0)=|\zeta_{\rm exact} -\zeta_{\rm appr}(u=0)| \sim 2.9 \times 10^{-10}$.
For all $0 < u \leq u_f$ we have a relative error $\Delta \zeta(u)/\zeta_{\rm exact} \leq 10^{-5}~ \%$.
For the sake of completeness of this section, and anticipating results of section \ref{section4}, we show
in Table \ref{table2} values of the net kick velocity $V_k$ for $\rho_0=21^{\circ}$, $\gamma=0.5$ and
several values of $\alpha$, for truncations $N_P=5$, $N_P=6$ and $N_P=7$, which also indicates that the
truncation $N_P=7$ is sufficiently accurate.
\par The numerical method is stable for long time runs so that we are able to reach the
final configuration of the remnant black hole when the gravitational wave emission ceases\cite{aranha11}.
We can thus examine physical phenomena in the nonlinear regime
where full numerical relativity simulations might present some difficulties due, for instance, to the
limited computational domain. In this sense the results of the evolution of the above data may
be considered as complementary to full numerical relativity simulations on describing the post-merger regime.
\par Our numerical work in the present paper contemplates the whole interval $\alpha=[0,1]$
with fixed $\rho_0=21^{\circ}$ and $\gamma=0.5$. For comparison purposes, results for other
values of $\rho_0$ and $\gamma$ are also considered. Exhaustive numerical experiments show that after a sufficiently
long time $u \sim u_f$ all modal coefficients of the Galerkin expansion become constant up to
twelve significant digits, corresponding to the final time of computation $u_{f}$.
At $u_f$ the gravitational wave emission is considered to effectively cease. In fact, all the modal coefficients are
zero at $u=u_f$ except $A_{0,0}$, $A_{0,1}$ and $A_{1,1}$. From the final constant modal coefficients we
reconstruct $K(u_f,\theta,\phi)$ that, in all cases, can be approximated as
{\small
\begin{eqnarray}
\label{eq17}
K(u_f,\theta,\phi) \simeq \frac{K_{f}}{\cosh \gamma_{f}+ (n_{1f} \sin \theta \cos \phi+n_{3f} \cos \theta) \sinh \gamma_{f}},~~
\end{eqnarray}}
where
{\small
\begin{eqnarray}
\label{eq17-i}
\nonumber
n_{3f}&=&\Big(1+\Big(2 A_{1,1}(u_f)/A_{1,0}(u_{f})\Big)^2\Big)^{-1/2},\\
n_{1f}&=&\Big(2 A_{1,1}(u_f)/A_{1,0}(u_{f})\Big)~n_{3f},\\
\nonumber
\gamma_f&=&\tanh^{-1}\Big(A_{1,0}(u_f)/n_{3f} A_{0,0}(u_f) \Big),\\
\nonumber
K_f &=& 2 \cosh \gamma_f/A_{0,0}(u_f).
\end{eqnarray}}
\noindent An alternative to evaluate the parameter $K_f$ is the use of the initial data in the conserved
quantity $\zeta(u)$. With the final parameters $(K_f,\gamma_f,n_{1f},n_{3f})$ obtained from the final
modal coefficients, we have in all cases that the rms error of Eq. (\ref{eq17}) is of the order of, or smaller than $10^{-12}$.
The final configuration corresponds then to a boosted Schwarzschild black hole
(cf. (\ref{eq4})) with a final velocity $ v_{f}={\rm tanh} \gamma_{f}$
along the direction determined by ${\bf{n}}_f=(n_{1f},0,n_{3f})$, and a final Bondi rest mass $m_0 K_{f}^{3}$.
In general $\gamma_{f}< \gamma$ and $K_f > 1$. The angle $\rho_f=\arccos~(n_{3f})$ defines the direction of
the remnant with respect to the $z$-axis. Within the numerical error of our
computation we have $(n_{1f})^2+(n_{3f})^2=1$, as expected.
\par The values of the parameters of the remnant black hole are one of the basic results
to be extracted from our numerical experiments. Also, by reconstructing $K(u,\theta,\phi)$ for all $u>u_0$, we can obtain
the time behavior of important physical quantities, as for instance the total impulse imparted to the merged
system by the emission of gravitational waves. Our numerical results are displayed in Table \ref{table3} and constitute
the basis of the analysis of the gravitational wave recoil, in the next Section.

\section{Gravitational wave momentum fluxes and the kick processes in a non-head-on collision\label{section4}}

We can now discuss the processes of momentum extraction of the system by constructing,
via the numerically integrated function $K(u,\theta,\phi)$, the time evolution of the net momentum fluxes
carried out by gravitational waves. In Fig. \ref{fig1} (top) we display the curves of the net momentum fluxes
$P_{W}^{x}(u)$ and $P_{W}^{z}(u)$ for mass ratios $\alpha=0.2,~0.5$. We see that
an initial regime of positive flux $P_{W}^{z}(u)$ is present with consequent increase of
the linear momentum along the $z$-axis. $P_{W}^{x}(u)$ is negative for all $u_0 < u \leq u_f$.
Therefore the main contribution to the associated impulses comes from the final dominant deceleration regime
until the system reaches the final configuration of the remnant black hole.
The initial regime of positive flux $P_{W}^{z}(u)$ is not present
in the case of large $\alpha$ and $\rho_0$.
\subsection{The integrated fluxes of gravitational waves and the total impulse imparted to the merged system}
\par Integrating in time the conservation Eq. (\ref{eq13}) we find that
{\small
\begin{eqnarray}
\label{eq18}
{\bf P}(u)- {\bf P}(u_0)={\bf I}_{W}(u),
\end{eqnarray}}
where ${\bf I}_{W}(u)=({I}_{W}^{x}(u),0,{I}_{W}^{z}(u))$
is the impulse imparted to the merged system by the gravitational waves emitted up to the time $u$.
In the above,
{\small
\begin{eqnarray}
\label{eq20-i}
{I}_{W}^{x}(u)=\int^{u}_{u_0} P_{W}^{x}(u)~d u,~~~{I}_{W}^{z}(u)=\int^{u}_{u_0} P_{W}^{z}(u)~d u,
\end{eqnarray}}
cf. Eqs. (\ref{eq15-i})-(\ref{eq15-ii}). The net total impulse imparted to the system has a dominant
contribution from the deceleration regime (when ${\bf{I}}_{W} < 0$) and will correspond
to a net kick on the merged system.
\begin{figure}
\begin{center}
{\includegraphics*[height=5.2cm,width=7.25cm]{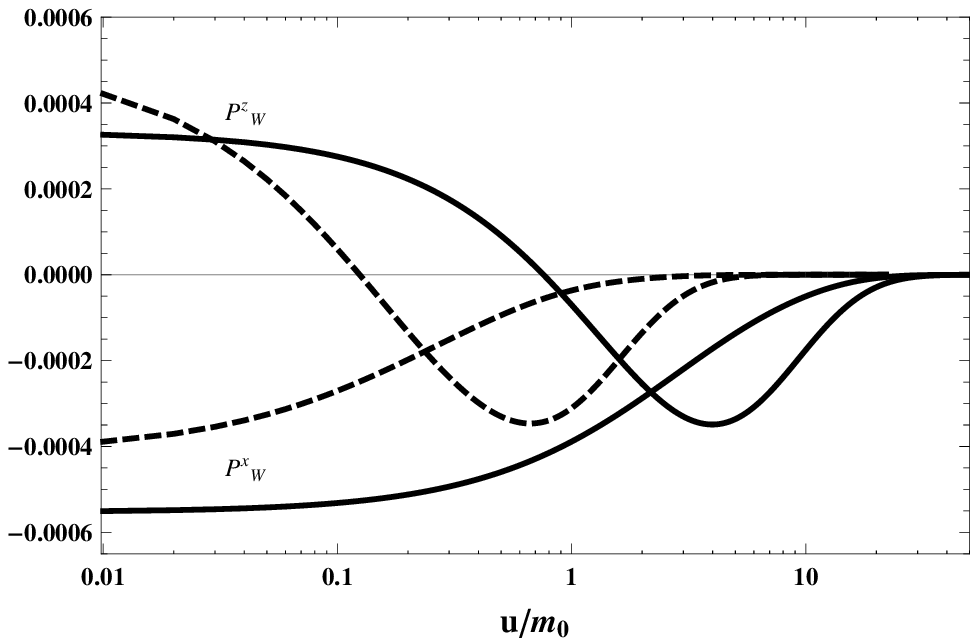}}
{\includegraphics*[height=5.2cm,width=7.5cm]{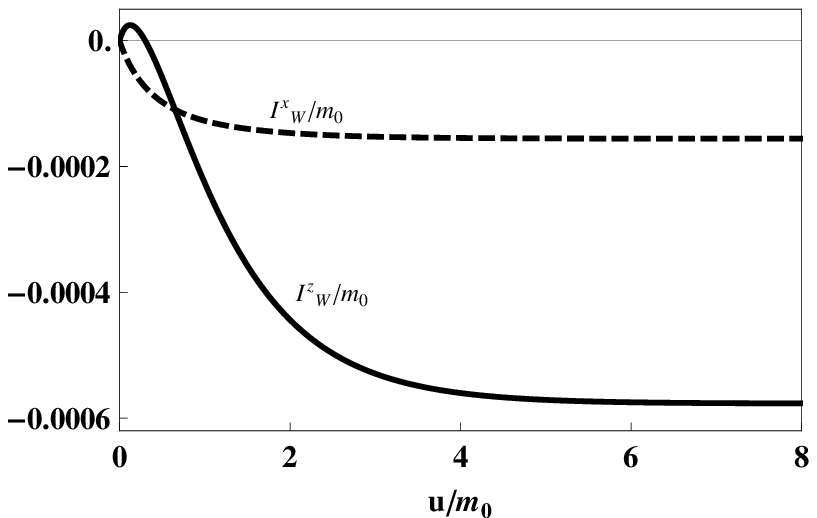}}
\caption{(top) Linear-log plot of the net momentum fluxes $P_{W}^{x}(u)$ and $P_{W}^{z}(u)$ for $\alpha=0.2$ (dashed curves) and
$\alpha=0.5$ (continuous curves). For both cases, an initial regime of positive
momentum flux along the $z$ axis is present. The main contribution to the
associated impulse comes from the final dominant deceleration phase. (bottom) Plot of the gravitational wave impulses
associated with the net momentum fluxes for $\alpha=0.2$ on top.
Both curves tend to a constant negative value, when the gravitational wave emission ceases.}
\label{fig1}
\end{center}
\end{figure}
\par The behavior of the impulse is illustrated in Fig. \ref{fig1} (bottom) for the case $\alpha=0.2$ of Fig. \ref{fig1} (top).
As expected the curve for $I_{W}^{z}(u)$ presents a local maximum, when the area measured below the curve of
$P_{W}^{z}(u)$ starts to give a negative contribution to the impulse. For large $u \sim u_f$ both curves tend
to a constant negative value, corresponding to the final configuration of the system.
Fig. \ref{fig1} (bottom) also illustrates the dominance of the deceleration regime in the processes.
The plateau is considered to be reached when $|{\bf I}_{W}(u)-{\bf I}_{W}(u+h)| \lesssim 10^{-10}$,
where $h$ is the stepsize of the integration used for the evaluation of ${\bf I}_{W}(u)$.
At this stage the remnant black hole has a momentum ${\bf P}_f \equiv {\bf P}(u_f)=(n_{1f},0,n_{3f})~P_f$, with
{\small
\begin{eqnarray}
\label{momentumF}
P_f=m_0 K_{f}^{3} \sinh \gamma_f,
\end{eqnarray}}
whose distribution as function of $\alpha$, for several $\rho_0$, is given in \cite{aranha11}.
Table \ref{table3} contains data characterizing ${P}_f/m_0$ in the case $\rho_0=21^{\circ}$.
\par From Eq. (\ref{eq18}) we can derive that
\small
\begin{eqnarray}
\label{eq21}
P^{x}(u_f)-P^{x}(u_0)=I^{x}_{W}(u_f),~~P^{z}(u_f)- P^{z}(u_0) =I^{z}_{W}(u_f),~~
\end{eqnarray}
where the right-hand sides of (\ref{eq21}) are the nonzero components of the net total impulse
${\bf I}_W(u_f)$.
We remark that for larger values of $\alpha$, the final impulse $I_{W}^{x}(u_f)$ is larger than $I_{W}^{z}(u_f)$,
contrary to the cases $\alpha=0.2,~0.5$. This is shown in Figs. \ref{fig3} where we plot
the net momentum fluxes $P_{W}^{x}(u)$ and $P_{W}^{z}(u)$, and the associated impulses
$I_{W}^{x}(u)$ and $I_{W}^{z}(u)$, for the case $\alpha=1$.
\par The purpose of Figs. \ref{fig3} is two-fold. First to show that, for high values of the
mass ratio, the initial phase of positive gravitational wave flux along the $z$-axis is absent, with an
overall deceleration of the merged system during the whole regime of gravitational wave emission,
typical of gravitational bremsstrahlung. Second, the total final impulse along the $z$ axis is about one order of magnitude smaller than the
total impulse along the $x$ axis; both of them are at least one order of magnitude larger than the corresponding
impulses for $\alpha=0.2$.
We should also note that in this case $\alpha=1$
(initial data corresponding to equal-mass black holes) the net gravitational wave flux and the
associated impulse are nonzero. This is not in contradiction with numerical relativity
results for binary black hole inspirals since only the case of a head-on collision
($\rho_0=0^{\circ}$ in our data (\ref{eq16})) can approximate, and therefore be compared with
the post-merger phase of black hole binary inspirals. Actually our initial data with $\rho_0=0^{\circ}$  yields a
zero net momentum flux for $\alpha=1$ as expected (cf. also \cite{aranha}).
In this sense the results of Table \ref{table3} (cf. also Fig. \ref{fitchett}) should be interpreted in the
light of gravitational configurations emitting gravitational waves other than black hole binary inspirals.
Since, to our knowledge, all black hole spacetimes examined in the literature of numerical relativity correspond either to
binary black hole inspirals or to head-on collisions we have presently no reference for comparison or analogy
with our results. Our best guess is that this system might be a candidate to an approximate description
of the post-merger phase of a non-head-on collision of black holes not preceded by a pre-merger
inspiral phase, as for instance colliding black holes in pre-merger unbounded trajectories
as schematically illustrated in Fig. \ref{figDD}.
Such a system reminds us of the near encounters of two bodies along unbounded trajectories that were
treated in Refs. \cite{NHC} where the authors, using a post-Newtonian treatment, showed the
universality of gravitational bremsstrahlung for these systems. However the use of black hole
initial conditions and distinct incidence angles, as well as a possible merger, were not contemplated
in their treatment due to the approximations used.
\begin{center}
\begin{table*}
\caption{Summary of our numerical results corresponding to an initial boost parameter
$\gamma=0.5$ and incidence angle $\rho_0=21^{\circ}$.}
\vspace{0.15cm}
{\small
\item[]
\begin{tabular}{|c|c|c|c|c|c|c|c|}
\hline
$\alpha$&$\eta$ & $K_f$ &$v_f/c=\tanh \gamma_f$ & $\rho_f$ & $-I_{W}^{x}(u_f)/m_0$ &  $- I_{W}^{z}(u_f)/m_0$ & $V_{k}~({\rm km/s})$   \\\hline
$0.025$ &$0.0238$ &$1.045113$  &$0.446425$&$0.37^{\circ}$ & $1.2792 \times 10^{-6}$ &$8.1563\times 10^{-6}$& $2.17$ \\ \hline
$0.050$ &$0.0453$ &$1.091454$ &$0.430866$ &$0.77^{\circ}$ & $5.6819 \times 10^{-6}$ &$3.3361\times 10^{-5}$ &  $7.80$ \\ \hline
$0.100$ &$0.0826$ & $1.187836$&$0.400255$& $1.65^{\circ}$ & $2.7635 \times 10^{-5}$&$1.3834 \times 10^{-4}$& $25.25$ \\ \hline
$0.150$ &$0.1134$ & $1.289162$ &$0.370480$&$2.64^{\circ}$ & $7.4349 \times 10^{-5} $& $3.1932 \times 10^{-4}$&  $45.91$ \\ \hline
$0.200$ &$0.1388$& $1.395446$&$0.341701$ &$3.77^{\circ}$ &$1.5575 \times 10^{-4}$ &  $5.7681 \times 10^{-4}$&  $65.96$  \\ \hline
$0.250$ &$0.1600$& $1.506701$&$0.314042$&$5.06^{\circ}$& $2.8314 \times 10^{-4}$&  $9.0763\times 10^{-4}$&  $83.39$  \\ \hline
$0.300$ &$0.1775$&$1.622933$& $0.287600$&$6.53^{\circ}$& $4.6910 \times 10^{-4}$ &  $1.3051 \times 10^{-3}$ &  $97.33$ \\ \hline
$0.400$ &$0.2040$& $1.870356$&$0.238634$&$10.17^{\circ}$& $1.0733 \times 10^{-3}$&  $2.2569 \times 10^{-3}$& $114.58$ \\ \hline
$0.500$ &$0.2222$& $2.137755$&$0.195219$&$15.06^{\circ}$& $2.0929 \times 10^{-3}$& $3.3174 \times 10^{-3}$& $120.45$  \\ \hline
$0.525$ & $0.2257$& $2.207669$&$0.185282$&$16.53^{\circ}$& $2.4287 \times 10^{-3}$& $3.5817 \times 10^{-3}$& $120.66$  \\ \hline
$0.600$ & $0.2344$& $2.425151$&$0.157731$&$21.77^{\circ}$& $3.6700 \times 10^{-3}$&  $4.3295 \times 10^{-3}$& $119.38$  \\ \hline
$0.700$ &$0.2422$& $2.732560$& $0.126756$&$31.14^{\circ}$& $5.9634 \times 10^{-3}$& $5.0986 \times 10^{-3}$ & $115.36$ \\ \hline
$0.800$ &$0.2469$ &$3.059988$&$0.103331$&$44.09^{\circ}$ &$9.1477 \times 10^{-3}$& $5.3968 \times 10^{-3}$ & $111.21$ \\ \hline
$0.900$ &$0.2493$ &$3.407441$& $0.088880$&$60.85^{\circ}$& $1.3413 \times 10^{-2}$&  $4.9651 \times 10^{-3}$& $108.45$ \\ \hline
$1.000$ &$0.2500$& $3.774920$&$0.084214$&$79.50^{\circ}$& $1.8965 \times 10^{-2}$&  $3.5148 \times 10^{-3}$&  $107.57$ \\ \hline
\end{tabular}}
\label{table3}
\end{table*}
\end{center}
\begin{figure}
\begin{center}
{\includegraphics*[height=5.2cm,width=7.25cm]{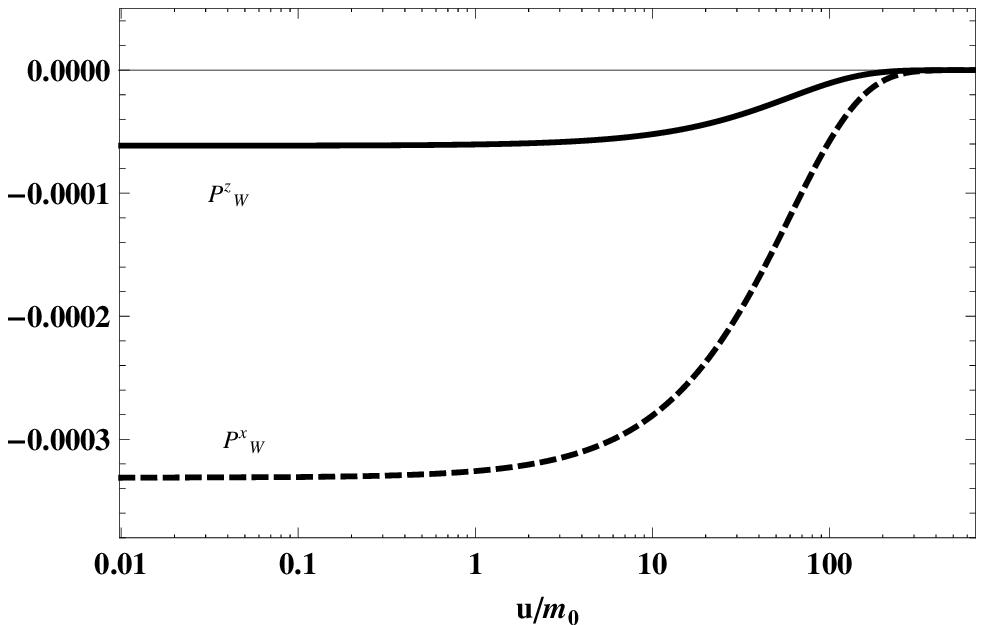}}\hspace{0.5cm}{\includegraphics*[height=4.92cm,width=7.25cm]{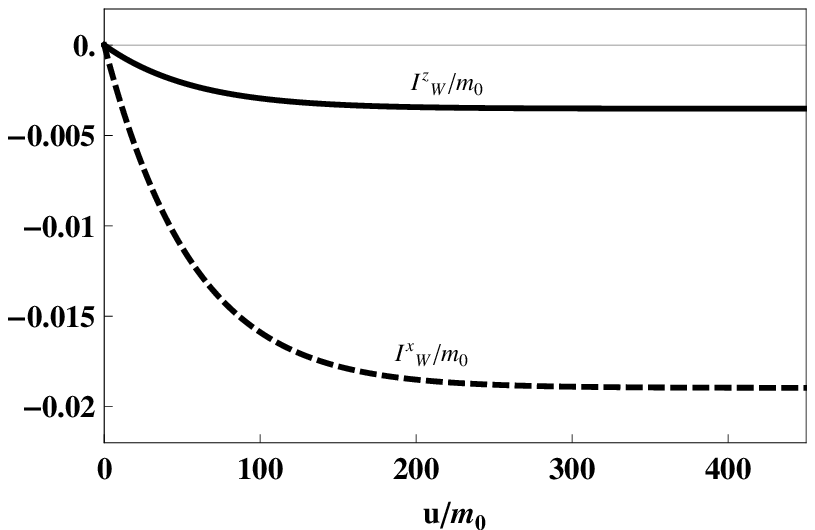}}
\caption{Linear-log plot of the net fluxes of momentum $P_{W}^{x}(u)$ and $P_{W}^{z}(u)$ (top)
and the associated total impulses $I_{W}^{x}(u)$ and $I_{W}^{z}(u)$ (bottom), for the case $\alpha=1$
and incidence angle $\rho_0=21^{\circ}$.
}
\label{fig3}
\end{center}
\end{figure}
\subsection{The total momentum and the kick velocity for non-head-on collisions}

\par We define the net kick velocity ${\bf V}_{k}$ of the merged system as proportional to the
momentum imparted to the system by the total impulse of gravitational waves. This definition
is based on the impulse function ${\bf I}_{W}(u)$ of the gravitational waves evaluated
at $u=u_f$ (cf. Eqs. (\ref{eq21})) and are in accordance with \cite{gonzalez}.
We obtain (restoring universal constants)
{\small
\begin{eqnarray}
\label{eq23-0}
{\bf V}_{k}&=&\frac{c}{m_0 K_{f}^{3}}~{\bf I}_{W}(u_f),
\end{eqnarray}}
with modulus
{\small
\begin{eqnarray}
\label{eq23}
V_{k}= \frac{c}{m_0 K_{f}^{3}}~\sqrt{I_{W}^{x}(u_f)^2+I_{W}^{z}(u_f)^2} ~,
\end{eqnarray}}
where $m_0 K_{f}^{3}$ is the rest mass of the remnant black hole.
Taking into account the momentum conservation equations (\ref{eq18}) evaluated at $u=u_f$
we interpret the definition (\ref{eq23-0}) as the balance between the Bondi momentum of
the system and the impulse of the gravitational waves in a zero-initial-Bondi-momentum frame,
which can then be compared with the results of the existing literature. We remark that the
zero-initial-Bondi-momentum frame is the inertial frame related to the asymptotic Lorentz frame
used in our computations by a velocity transformation with velocity parameter
${\small {\bf v}_B={\bf P} (u_0)/m_0 K_f^3}$~; in the parameter domain of our
numerical experiments the relativistic corrections in this transformation may be neglected.
\par We evaluate $V_{k}$ for several values of $\alpha$, for fixed $\gamma=0.5$ and
incidence angle $\rho_0=21^{\circ}$. The results are summarized in Table \ref{table3} where
we use the symmetric mass parameter $\eta=\alpha/(1+\alpha)^2$. In Fig. \ref{fitchett} we plot the points
$(V_{k},\eta)$ from Table \ref{table3}. The continuous curve is the least-square-fit of the points to the
empirical analytical formula,
{\small
\begin{eqnarray}
\label{eq24}
V= A \eta^2 (1- 4 C \eta)^{1/2} (1+B \eta) \times 10^{3} ~{\rm km/s},
\end{eqnarray}}
with best fit parameters $A\simeq 3.63538199$, $B\simeq 2.46152380$ and $C \simeq 0.91379025$.
The maximum net antikick obtained is $\simeq 120.66 ~{\rm km/s}$ at $\eta \simeq 0.2257$.
\begin{figure}
\begin{center}
{\includegraphics*[height=5.6cm,width=8.5cm]{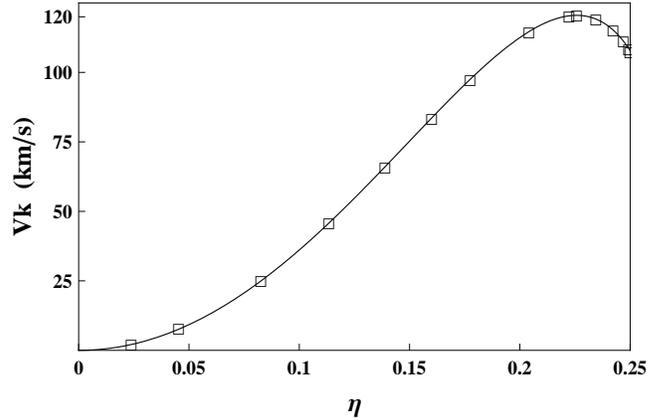}}
\caption{Plot of the points ($V_{k},\eta$) given in Table \ref{table3}, for $\rho_0=21^{\circ}$.
The continuous curve is the best fit of the points to the empirical law (\ref{eq24}).
The maximum of the curve corresponds to $V_{k} \simeq 120.66~ {\rm km/s}$ for $\eta \simeq 0.2257$
($\alpha \simeq 0.5249$). The normalized rms error of the fit is $\simeq 0.13 \% $. The nonzero kick for
$\alpha=1$ confirms that the non-head-on initial data have no connection with black hole binary inspirals.}
\label{fitchett}
\end{center}
\end{figure}
The parameter $C$ is empirically introduced to account for the nonzero kick velocity in the non-head-on
case  with mass ratio $\alpha=1$. We note that (\ref{eq24}) reduces to the Fitchett law\cite{fitchett,blanchet1}
for $C=1$. We must emphasize that the results shown in Fig. \ref{fitchett} have
no connection with black hole binary inspirals, as discussed already, rather possibly with two
colliding black holes in pre-merger unbounded trajectories as schematically shown in Fig. \ref{figDD}.
We note however that our data (\ref{eq16}) for $\rho_0=0^{\circ}$ (head-on case) yield a Fitchett distribution,
as expected and discussed in the next subsection.
\par The nonzero kick velocity for the non-head-on data with $\alpha=1$ and $\rho_0=21^{\circ}$ deserves
a more detailed discussion.
In this case we can evaluate the components of the initial Bondi-Sachs momentum to be,
$P^{x}(0)/m_0 \simeq 4.489093$, $P^{z}(0)/m_0 \simeq 0.832004$ and $P^{y}(0)/m_0 \simeq 0$,
with respect to an asymptotic Lorentz observer. This momentum vector, which lies in the right quadrant of the upper hemisphere $z>0$
of the plane $x-z$, makes an angle $\Theta_B=\arctan |P^{x}(0)/P^{z}(0)| \simeq 1.387537$ radians (or $\Theta_B \simeq 79.5^{\circ}$)
with the positive $z$-axis. This is also the direction of the nonzero momentum of the remnant with respect to the same asymptotic
Lorentz frame, determined by the angle $\rho_f$, which satisfies $\rho_f=(180^{\circ}-\rho_0)/2$ for $\alpha=1$ and
any $\rho_0$ (cf. Table \ref{table3} and Ref. \cite{aranha11}). The axis determined by $\rho_f \equiv \Theta_B$
actually plays an important role in the dynamics. If we take a new frame with its $z$-axis coinciding
with this axis the net gravitational wave momentum flux vector ${\bf P}_{W}(u)$ lies along the new $z$-axis
for all $u$. This was verified numerically by evaluating the ratios of the computed fluxes, used to
construct Fig. \ref{fig3} (right), sampled in the interval $0 < u/m_0 \leq 890$, yielding in all cases
$\arctan|I_{W}^{x}(u)/I_{W}^{z}(u)| \simeq 1.387541$ with a relative error of the order of $10^{-6}$.
Still in this new frame the data will not be symmetric under $\theta \rightarrow \pi - \theta$,
leading to a nonzero net gravitational wave momentum flux, contrary to the case of merging binary
inspirals and head-on collisions. As expected the $z$-axis of the new frame
is the direction of the kick velocity since $\arctan |I_{W}^{x}(u_f)/I_{W}^{z}(u_f)| \simeq 1.387547 ~{\rm rad}$
or $\simeq 79.5^{\circ}$ (within the precision of data in Table \ref{table3}).
Finally in the evolution of these data we had no evidence of the presence of a {\it junk}
radiation component, as discussed already. Therefore it is unlikely that the nonzero net gravitational wave momentum
flux in this case is a contribution of a junk radiation component.
\begin{figure}[t]
\begin{center}
{\includegraphics*[height=5.5cm,width=8.5cm]{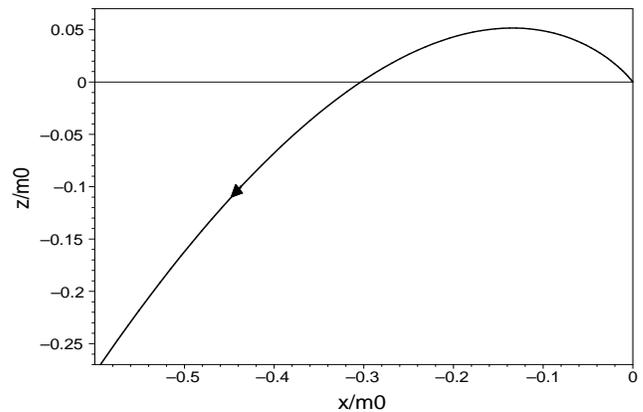}}
\caption{Plot of the integral curve ${\bf x}(u)$ of the wave impulse vector field ${\bf I}_{W}(u)$
which can give a schematic picture of the motion of the system in the zero-momentum-Bondi frame, for $\alpha=0.2$
and $\rho_0=21^{\circ}$. The asymptote of the curve as $u \rightarrow u_f$ makes an angle $\Theta_f \simeq 15.11^{o}$
with the negative $z$-axis of this frame.}
\label{figCM}
\end{center}
\end{figure}
\par In general, in a zero-initial-Bondi-momentum frame, the Bondi momentum of the merged system
satisfies ${\bf P}(u)={\bf I}_{W}(u)$ so that an integral curve ${\bf x}(u)$ of the wave impulse
vector field ${\bf I}_{W}(u)$, defined as $d{\bf x}/du={\bf P}(u)$, can give a schematic
picture of the motion of the merged system in this frame. In Fig. \ref{figCM} we
display this integral curve for $\alpha=0.2$ and $\rho_0=21^{\circ}$, generated with initial
conditions $x(u_0)=0=z(u_0)$ in the zero-initial-Bondi-momentum frame.
The initial phase of positive momentum flux along $z$ is responsible for the curved form
of the trajectory in the semiplane $z > 0$. For $u \rightarrow u_f$ the curve approaches the asymptote with
angle $\Theta_f=\arctan (I_{W}^{x}(u_f)/I_{W}^{z}(u_f)) \simeq 15.11^{\circ}$ with respect
to the negative $z$-axis of the zero-initial-Bondi-momentum frame, which is actually the
direction of the kick velocity in this frame.
In the case $\alpha=1$ the integral curve is a straight
line with angle $\Theta_f \simeq 79.5^{\circ} \equiv (180^{\circ}-21^{\circ})/2$ with respect to the $z$-axis
of the zero-initial-Bondi-momentum frame (cf. Table \ref{table3}).

\subsection{Head-on kick distributions: a comparison to results of the literature}

We now compare and discuss our evaluations of kick velocities
for the head-on case ($\rho_0=0^{\circ}$) with the results
of the numerical relativity simulations (bridged with analytical approximations)
of Refs. \cite{blanchet} (PN+CLA) and \cite{sopuerta} (CLA), and with results of full numerical
relativity of Refs. \cite{baker,gonzalez,gonzalez1,camp4,camp2}. These works treat
black hole inspiral binary systems and are related to our results in the sense that
head-on collisions can be an approximation model to the final plunge
to merger in binary inspirals. The kick distribution obtained in our simulations satisfies the Fitchett law
(cf. also \cite{aranha}), so that the recoil evaluated from the BS net momentum fluxes
can give a complementary information on the momentum extraction, contributing to the
understanding of the processes of producing kicks and anti-kicks. The post-merger phase
corresponding to our head-on initial data exhibits two distinct regimes,
an initial positive, followed by a dominant negative phase of net momentum flux,
a pattern consistent with kick and antikick contributions.
\par For our numerical experiments in this subsection we have adopted an
initial data with boost parameter $\gamma=0.56$, chosen to match the
maximum kick velocity of \cite{blanchet}.
The results are displayed in Fig. \ref{ALL} together with the main results
of Refs. \cite{baker,sopuerta,gonzalez,gonzalez1,camp4,camp2,blanchet}. The points $(V_k,\eta)$ of our simulations
are fitted by the Fitchett law $V_k=A \eta^2 (1- 4 \eta)^{1/2} (1+B \eta) \times 10^{3} ~{\rm km/s}$, with best fit parameters
$A=5.4719439$ and $B=3.874397$. The normalized rms of the fit to the points is $\simeq 0.36\%$.
The results of the PN+CLA\cite{blanchet} are
shown by their Fitchett curve, fitted with parameters $A=9.5$ and $B=0.3$, and four points with error estimates.
The CLA treatment of \cite{sopuerta} is given by the curve
$V_k=11.292 \eta^2 (1- 4 \eta)^{1/2} (1-1.6070 \eta+1.9076 \eta^2) \times 10^{3} ~{\rm km/s}$.
The full numerical relativity calculations of Gonz\'alez et al.\cite{gonzalez} are given by the
Fitchett curve fitted with parameters $A=12$, $B=-0.93$, with a maximum kick $V_k=175.2 \pm 11 {\rm km/s}$
at $\eta= 0.195 \pm 0.005$.
The agreement of our numerical results with the corresponding points of numerical relativity simulations
is to $5\% -8\%$ for the range of mass ratios $0.3 \leq \alpha \leq 0.95$.
Larger discrepancies occur for smaller mass ratios, of the order of $15\%-31\%$.
These systematic discrepancies could possibly be due to a missing inspiral contribution since
the RT dynamics covers only the post-merger phase of the system.
However, in spite of these discrepancies, our results confirm that the
main contribution to the kicks comes from the post merger phase.
\par A full numerical relativity treatment of the gravitational wave recoil in head-on collisions of two black holes
was done in Choi et al.\cite{choi}, where both the black hole mass ratio and spins are varied.
For the unequal mass case ($\alpha=0.667$) without spin they obtained a net kick velocity $V_k \sim 2.76 ~{\rm km/s}$,
for a large initial separation ($L/M=12.24$) with the black holes starting at rest. The latter two conditions
of the initial configuration can possibly be the reason for the discrepancy of two orders of magnitude
with our result for the same mass ratio, which is $V_k \sim 121 ~{\rm km/s}$ (for $\gamma=0.56$)
and $V_k \sim 70~{\rm km/s}$ (for $\gamma=0.5$). In fact we have evaluated the net kick for a sufficiently
small value $\gamma=0.26$ and the same mass ratio $\alpha=0.667$ in our data.
We obtained $V_k \simeq 2.59 ~{\rm km/s}$,
of the order of their reported value. The paper \cite{choi} does not provide values of kicks
for other mass ratios, which would be a valuable comparison to our results in the BS approach.
\begin{figure}
\begin{center}
\hspace{-1.2cm}
{\includegraphics*[height=5.8cm,width=9.5cm]{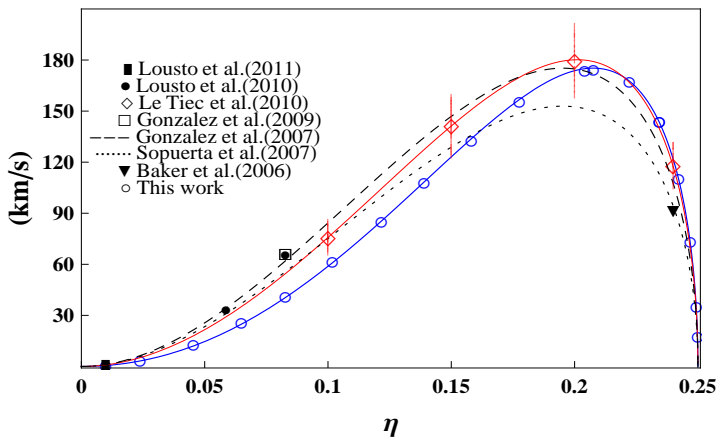}}
\caption{Comparing our numerical results for kicks generated in the post-merger phase of a head-on collision
with CLA\cite{sopuerta} and PN+CLA\cite{blanchet} results, and with full numerical
relativity results\cite{baker,gonzalez,gonzalez1,camp4,camp2} for binary black hole mergers.}
\label{ALL}
\end{center}
\end{figure}
\par Finally we refer to the works of Shibata et al.\cite{shibata} and Sperhake et al.\cite{shibata1}
examined the collision of two equal black holes, with anti-parallel initial velocities,
directed at each other with an impact parameter. We consider that the initial data used by
the authors do not actually describe the same physical situation of our non-head-on data.
In our case, for any value of the incidence angle $\rho_0$, a black hole
is formed while in those authors' data this depends on the impact parameter, with a critical threshold
for the formation of black holes. Furthermore they report a huge increase in the efficiency of energy extraction
by gravitational waves as compared to the case of zero impact parameter (pure head-on). In contrast,
in our system, head-on collisions constitute an upper bound for the efficiency. The efficiency decreases monotonically
with the increase of $\rho_0$ for any $\alpha$, as examined in Ref. \cite{aranha11}. Therefore there appears to
be no obvious connection between the impact parameter $b$ of their data and the parameter $\rho_0$ of our initial data.

\section{Final Discussions and Conclusions}

The present paper extends our work in \cite{aranha11} by examining the
gravitational wave recoil in non-axisymmetric Robinson-Trautman spacetimes
in the BS characteristic formalism. The initial data already have
a common apparent horizon so that the RT dynamics actually evolves a distorted black hole,
reminiscent of the CLA approach\cite{zerilli,teuk}. The characteristic initial data
used in the RT dynamics were derived and interpreted in \cite{aranha11} as corresponding
to the early post-merger state of two boosted colliding black holes with a common apparent horizon.
The basic parameters of the initial data are the mass ratio $\alpha$, the boost parameter $\gamma$ and
the incidence angle $\rho_0$. Head-on collisions correspond to the particular case $\rho_0=0^{\circ}$ and
only in this case can our results be compared with numerical relativity calculations for
binary black hole spacetimes and head-on collisions. Our analysis is based on the Bondi-Sachs momentum
conservation laws that regulate the radiative transfer processes of the system. We use a numerical code
based on the Galerkin spectral method\cite{aranha11}, which is sufficiently accurate and stable for long time runs,
and we evaluate the parameters of the remnant black hole, the gravitational
wave net momentum fluxes and the impulse imparted to the system by the emitted radiation.
\par In the paper we restricted our numerical computations to two distinct sets of initial data:
(i) initial data corresponding to a non-head-on collision with $\rho_0=21^{\circ}$, and $\gamma=0.5$ fixed;
(ii) initial data corresponding to head-on collisions ($\rho_0=0^{\circ}$ ) and $\gamma=0.56,~0.26$,
the values of $\gamma$ being adopted in order to compare our simulations with numerical relativity simulations
of inspiral black hole binaries.
A feature to be remarked in the case of non-head-on initial data (i) is the nonzero net gravitational wave flux
for equal mass black holes, contrary to the case of head-on collisions and inspiral black hole binaries,
showing that the data (i) has no connection or analogy with black hole binary inspirals and head-on collisions
examined in numerical relativity simulations. We suggest that these data might be a candidate to an approximate
description of a post-merger phase of a non-head-on collision of black holes in unbounded trajectories.
The numerical results for the head-on initial data (ii) were compared
with published results of kick distributions in numerical relativity, as
from PN+CLA\cite{sopuerta,blanchet} to numerical relativity evaluations\cite{baker,gonzalez,gonzalez1,camp4,camp2}
of kicks in the collapse and merger of nonspinning black hole binaries.
This comparison is made taking into account that
a head-on configuration may be seen as an approximation to the final plunge to merger
of inspiral binaries. For the range of mass ratios $(0.3 \leq \alpha \leq 0.95)$ we have an agreement
to $5\% - 8\%$, but larger discrepancies occur
for smaller mass ratios. Our results confirm that the post merger phase gives a substantial
contribution to the kicks.
\par The full numerical relativity experiment of Choi et al.\cite{choi}
for head-on collisions of nonspinning black holes obtained a kick velocity
two orders of magnitude smaller then the ones reported in our results shown in Fig. \ref{ALL}.
We evaluated the net kick velocity using the same mass ratio but a
smaller boost parameter to approach their initial data that used black holes starting at rest.
We obtained a kick of the same order.
Finally we discussed Refs. \cite{shibata,shibata1}, where the collision of two equal-mass black
holes with an impact parameter is examined, and compared with our non-head-on case.
The authors report a huge increase in the efficiency of energy extraction
as compared to the case of zero impact parameter (pure head-on). In contrast,
in our system, head-on collisions $(\rho_0=0^{\circ})$ constitute an upper bound for the efficiency,
which decreases monotonically with the increase of $\rho_0$\cite{aranha11}.
Therefore there appears to be no obvious connection between the impact parameter of their data and the
parameter $\rho_0$ of our initial data.
\section*{Acknowledgements}
{The authors acknowledge the partial financial support of CNPq/MCTI-Brazil, through a Post-Doctoral Grant (RFA),
Research Grant (IDS), and of FAPES-ES-Brazil (EVT). RFA acknowledges the hospitality and partial financial
support of the Center for Relativistic Astrophysics, Georgia Institute of Technology, Atlanta, GA, USA.
We thank Dr. Alexandre Le Tiec for kindly providing us with
results of Ref. \cite{blanchet}, included in Fig. \ref{ALL} of this paper.


\begin{thebibliography}{99}

\bibitem{pretorius} F. Pretorius, {\it Physics of Relativistic Objects in Compact Binaries: from Birth to
Coalescence}, Eds. Colpi M, Casella P, Gorini V, Moschella U and Possenti A, (Astrophysics and Space Science Library Series, Vol. 359, Springer, Heidelberg), p. 305 (2009).

\bibitem{baker} J. B. Baker, J. Centrella, D. Choi, M. Koppitz, M. van Meter and M. C. Miller, {\it Astrophys. J.} {\bf 653} L93-L96 (2006).

\bibitem{merritt} D. Merrit, M. Milosavljevi\'c, M. Favata, S. S. Hughes and D. E. Holz, {\it Astrophys. J.} {\bf 607} L9 (2004).

\bibitem{favata} M. Favata M, S. A. Hughes and D. E. Holz, {\it Astrophys. J.} {\bf 607} L5 (2004).

\bibitem{komossa} S. Komossa, H. Zhou and H. Lu , {\it Astrophys. J.} {\bf 678} L81 (2008); P. G. Jonker, M. A. P. Torres, A. C. Fabian,
M. Heida, G. Miniutti and D. Pooley, {\it Mon. Not. R. Astron. Soc} {\bf 407} 645 (2010).

\bibitem{blanchet1} L. Blanchet, M. S. S. Qusailah and C. M. Will, {\it Astrophys. J.} {\bf 635} 508 (2005).

\bibitem{sopuerta}  C. F. Sopuerta, N. Yunes and P. Laguna, {\it Phys. Rev.} D {\bf 74} 124010 (2006);
Erratum: {\it Phys. Rev.} D {\bf 75} 069903(E) (2007).

\bibitem{gonzalez0} J. A. Gonz\'alez, M. Hannam, U. Sperhake, B. Br\"ugmann and S. Husa,
{\it Phys. Rev. Lett.} {\bf 98} 091101 (2007).

\bibitem{camp1} M. Campanelli, C. O. Lousto, Y. Zlochower and D. Merritt, {\it Astrophys. J.} {\bf 659} L5-L8 (2007).

\bibitem{gonzalez} J. A. Gonz\'alez, U. Sperhake, B. Br\"ugmann, M. Hannam and S. Husa, {\it Phys. Rev. Lett.} {\bf 98} 091101 (2007).

\bibitem{gonzalez1} J. A. Gonz\'alez, U. Sperhake and B. Br\"ugmann, {\it Phys. Rev.} D {\bf 79} 124006 (2009).

\bibitem{camp3} C. O. Lousto, N. Nakano, Y. Zlochower Y and M. Campanelli, {\it Phys. Rev. Lett.} {\bf 104} 211101 (2010).

\bibitem{camp4} N. Nakano, Y. Zlochower, C. O. Lousto and M. Campanelli, {\it Phys. Rev.} D {\bf 84} 124006 (2011).

\bibitem{camp2} C. O. Lousto Y. Zlochower, {\it Phys. Rev. Lett.} {\bf 106} 041101 (2011).

\bibitem{blanchet} A. Le Tiec, L. Blanchet and C. M. Will, {\it Class. Q. Grav.} {\bf 27} 012001 (2010).

\bibitem{choi} D. Choi, B. J. Kelly, W. D. Boggs, J. G. Baker, J. Centrella and J. van Meter {\it Phys. Rev.} D {\bf 76} 104026 (2007).

\bibitem{rezzolla} L. Rezzolla, R. P. Macedo and J. L. Jaramillo, {\it Phys. Rev. Lett.} {\bf 104} 221101 (2010).

\bibitem{rezzolla1} J. L. Jaramillo, R. P. Macedo, P. Moesta and L. Rezzolla, {\it Phys. Rev.} D {\bf 85} 084030 (2012).

\bibitem{aranha1} R. F. Aranha, I. Dami\~ao Soares and E. V. Tonini, {\it Phys. Rev.} D {\bf 81} 104005 (2010).

\bibitem{aranhaT} R. F. Aranha, H. P. Oliveira, I. Dami\~ao Soares and E. V. Tonini, {\it Int. J. Mod. Phys.} D {\bf 17} 2049 (2008).

\bibitem{rt} I. Robinson and A. Trautman, {\it Phys. Rev. Lett.} {\bf 4} 431 (1960); I. Robinson and A. Trautman, {\it Proc. Roy. Soc. London} A {\bf 265} 463 (1962).

\bibitem{bondi} H. Bondi H, M. G. J. van der Berg and A. W. K. Metzner, {\it Proc. Roy. Soc. London} A {\bf 269} 21 (1962).

\bibitem{sachs} R. K. Sachs, {\it Proc. Roy. Soc. London} A {\bf 270} 103 (1962); R. K. Sachs, {\it J. Math. Phys.} {\bf 3} 908 (1962).

\bibitem{aranha11} R. F. Aranha, I. Dami\~ao Soares and E. V. Tonini, {\it Phys. Rev.} D {\bf 85} 024003 (2012).

\bibitem{zerilli} F. Zerilli, {\it Phys. Rev. Lett.} {\bf 24} 737 (1970); C. T. Cunningham,
R. H. Price and V. Moncrief, {\it Astrophys. J.} {\bf 224} 643 (1978).

\bibitem{teuk} S. A. Teukolsky, {\it Phys. Rev. Lett.} {\bf 29} 1114 (1972).

\bibitem{chrusciel} P. Chrusciel, {\it Commun. Math. Phys.} {\bf 137} 289 (1991); P. Chru\'sciel P
and D. B. Singleton, {\it Commun. Math. Phys.} {\bf 147} 137 (1992).

\bibitem{AH} K. P. Tod, {\it Class. Quantum Grav.} {\bf 6} 1159 (1989); H. P. Oliveira, E. Rodrigues and I. Dami\~ao Soares,
{\it Braz. J. Phys.} {\bf 41} 314 (2011).

\bibitem{sachs1} R. K. Sachs, {\it Phys. Rev.} {\bf 128} 2851 (1962).

\bibitem{cqgIvano} R. F. Aranha, I. Dami\~ao Soares and E. V. Tonini, {\it Class. Quantum Grav.} {\bf 30} 025014 (2013).

\bibitem{fletcher} C. A. Fletcher, {\it Computational Galerkin Methods} (Berlin: Springer-Verlag, 1984).

\bibitem{aranha}  R. F. Aranha, I. Dami\~ao Soares and E. V. Tonini, {\it Phys. Rev.} D {\bf 82} 104033 (2010).

\bibitem{NHC}
P. C. Peters, {\it Phys. Rev.} D {\bf 1} 1557 (1970); M. Turner, {\it Astrophys. J.} {\bf 216} 610 (1977);
S. D. Kov\'acs and K. S. Thorne, {\it Astrophys. J.} {\bf 217} 252 (1977); M. Turner and C. W. Will, {\it Astrophys. J.} {\bf 220} 1107 (1978);
S. D. Kov\'acs and K. S. Thorne, {\it Astrophys. J.} {\bf 224} 62 (1978).

\bibitem{fitchett} M. J. Fitchett, {\it Mon. Not. R. Astron. Soc.} {\bf 203} 1049 (1983).

\bibitem{shibata} M. Shibata, H. Okawa and T. Yamamoto, {\it Phys. Rev.} D {\bf 78} 101501 (2008).

\bibitem{shibata1} U. Sperhake, V. Cardoso, F. Pretorius, E. Berti, T. Hinderer and N. Yunes, {\it Phys. Rev. Lett.} {\bf 103} 131102 (2009).

\end{thebibliography}
\end{document}